# Possible way to achieve anomalous valley Hall effect by tunable intrinsic piezoelectric polarization in FeO$_2$SiGeN$_2$ monolayer

*Jianke Tian,[a] Jia Li,[\*,a,b] Hengbo Liu,[a] Yan Li,[a] Ze Liu,[a] Linyang Li,[a] Jun Li,[a] Guodong Liu,[a] Junjie Shi,[c]*

[a]*School of Science, Hebei University of Technology, Tianjin 300401, People's Republic of China*

[b]*College of Science, Civil Aviation University of China, Tianjin 300300, People's Republic of China*

[c]*State Key Laboratory for Artificial Microstructures and Mesoscopic Physics, School of Physics, Peking University Yangtze Delta Institute of Optoelectronics, Peking University, 5 Yiheyuan Street, Beijing, 100871, People's Republic of China.*

## ABSTRACT

Valley-related multiple Hall effect and piezoelectric response are novel transport characteristics in low-dimensional systems, however few studies have reported their coexistence in a single system as well as their coupling relationships. By first-principles calculations, we propose a multifunctional Janus semiconductor, i.e. FeO$_2$SiGeN$_2$ monolayer with large valley polarization of about 120 meV and in-plane piezoelectric polarization with $d_{11}$ of -0.71 − 4.03 pm/V. The magnetic anisotropy energy can be significantly regulated by electronic correlation strength and strain, which can be attributed to the change of competition relationship about Fe-3*d*-resolved magnetic anisotropy energy brought about by external regulatory means.

---
\* Corresponding author. *E-mail address*: lijia@cauc.edu.cn (J. Li).

Electronic correlation strength can induce phase transitions in Janus $FeO_2SiGeN_2$ monolayer from ferrovalley to quantum anomalous Hall phase, while the half-valley metallic state as the boundary of the phase transition can gererate 100% spin- and valley polarization. The related phase transition mechanism is analyzed based on the two-band strained $\boldsymbol{k} \cdot \boldsymbol{p}$ model. The presence of piezoelectric strain coefficients $d_{11}$ in valleytronic material makes the coupling between charge degrees of freedom and valley degrees of freedom possible, and the intrinsic electric field caused by the in-plane piezoelectric response provide the way to realize piezoelectric anomalous valley Hall effect. This work may pave a way to find a new member of materials with valley-related multiple Hall effect and stimulate further experimental works related to valleytronics and piezotronics.

## 1. Introduction

The Hall effect originates from the complex interaction between carriers driven by in-plane electric field and magnetic field, which reflects the significant transport in crystalline solids.[1-4] Berry curvature, which is regarded as the pseudomagnetic field, describes the geometric properties of the wave function. In general, the time-reversal (*T*) and inversion (*P*) symmetry breaking in crystalline solids can lead to a non-zero pseudomagnetic field (Berry curvature) to induce interesting Hall effects.[5-8] Under the *T*-broken systems, the non-zero Berry curvature coupled with spin degree of freedom may form the spin-polarized dissipationless chiral edge channels in quantum anomalous Hall effect (QAHE), and the Chern number *C* associated with quantum Hall conductance ($\sigma_{xy} = C\frac{e^2}{h}$) can be obtained by integrating the Berry curvature in a closed loop.[9-11] Meanwhile, the anomalous valley Hall effect (AVHE) is realized under the coupling of unequal Berry curvature and valley degree of freedom in the ferrovalley systems of *P*-broken.[12-15] For valleytronics, the existence of polarized-valleys (K and K' valleys), which is closely related to Berry curvature, provides a new way to regulate Bloch electrons resulting in some novel Hall effects, such as topology-engineered orbital Hall effect (OHE), quantum anomalous valley Hall effect

(QAVHE) and layer Hall effect (LHE).[16-18] Similar to valley polarization, piezoelectric polarization in piezotronics reflects the response of charge degree of freedom to external mechanical stress under *P*-broken. Overall, the symmetry breaking results in the disappearance of relevant invariant in the system, which has a profound effect on the transport properties of the relevant charge, spin and valley degrees of freedom.

In low-dimensional systems with remarkable quantum effects, interesting valley and piezoelectric coupling phenomena such as unconventional piezoelectricity, valley piezoelectric mechanism and Berry-Curvature Dopole have been investigated.[19-21] These quantum characteristics provide the new ideas for the preparation of multifunctional nanodevices, however there are still few explorations of valley and piezoelectric transport. The existence of P-broken makes valley polarization and piezoelectric polarization often coexist in a single system, which leads to a natural question that whether these novel valley Hall-related effect and piezoelectric response could be integrated in the ferrovalley materials.[22,23] Recently, the septuple-atomic-layer $MSi_2N_4$ (M=Mo, W) has been experimentally synthesized by chemical vapor deposition (CVD), which directly led to the prosperity of $MA_2Z_4$ (M=Mo, W, V, Nb, Ta, Ti, Cr, Hf, A=Si, Ge, Z=N, P, As) family with topological, superconducting, valley and piezoelectric properties.[18,24,25] However, the intrinsic long-range ferromagnetic sequence and valley-related multiple Hall effect (AVHE and QAVHE) have only been predicted in two-dimensional (2D) V-based systems (such as $VSiGeN_4$ and $VSi_2P_4$), which limit the application of topologically nontrivial ferrovalley materials base on the $MA_2Z_4$ family.[18,26] Therefore, expanding topologically nontrivial ferrovalley materials and integrating valley Hall-related effect and piezoelectric transport into such septuple atomic layer structures can facilitate the further experimental works of valley-related physics. Inspired by the successful synthesis of 2D $MSi_2N_4$ (M=Mo, W) without natural bulk counterparts, we here design the $FeO_2SiGeN_2$ monolayer and believe that it can be prepared experimentally in the future.

In this work, using first-principles calculations, we predicted a 2D multifunctional $FeO_2SiGeN_2$ monolayer with large valley and piezoelectric polarization. Biaxial strain and strong correlation effects ($U_{eff}$) can significantly regulate magnetic anisotropy energy (MAE), which is attributed to the competition of orbital-resolved MAE contributed by Fe atoms. The topological phase transitions, i.e. from FV to QAVH phase, can be induced by strain and $U_{eff}$, and a two-band strained $\mathbf{k} \cdot \mathbf{p}$ model is established to understand the key role of strain and $U_{eff}$ in phase transitions. Finally, we theoretically couples AVHE and piezoelectric polarization in $FeO_2SiGeN_2$ monolayer, and we speculate that this coupling mechanism is universal in 2D ferrovalley materials.

## 2. Computational methods

All calculations are performed using the projector augmented wave (PAW) method through the Vienna ab initio simulation package (VASP).[27-29] For exchange and correlation interactions, the generalized gradient approximation (GGA) of Perdew–Burke-Ernz-erhof functional (PBE) is treated.[30] To attain reliable results, the kinetic energy cutoff is set to be 500 eV and a 20 Å vacuum layer in the z direction is used to eliminate interactions between adjacent layers. The k meshes of Brillouin zone is adopted using Γ-centered 11×11×1 and 6×10×1 for calculations for structural relaxation and piezoelectric stress tensors $e_{ij}$, respectively. Atomic positions are fully relaxed until the maximum force on each atom is less than 0.01 eV/Å and total energy convergence criterion is set to be $10^{-6}$ eV. The elastic stiffness and piezoelectric stress tensors ($C_{ij}$ and $e_{ik}$) are calculated by using strain–stress relationship (SSR) and DFPT method, respectively. The $U_{eff}$ for the Fe-3$d$ electrons are handled via the PBE+U scheme for the electronic structure calculations.[31] The Spin-orbit-coupling (SOC) effect is included in the calculations and the VASP data are processed by VASPKIT code.[32,33] The phonon dispersion is based on a 4×4×1 supercell by using the PHONOPY code.[34] Ab initio molecular dynamic (AIMD) simulations adopt the NVT ensemble based on the Nosé-Hothermostat at 300 and 600K with a total of 6.0 ps.[35] The Berry curvatures of materials is calculated using maximally localized Wannier functions implemented in the WANNIER90, and the topological features of edge

states were investigated by using package WannierTools.[36,37]

## 3. Results and discussion

### 3.1. Atomic structures and stability

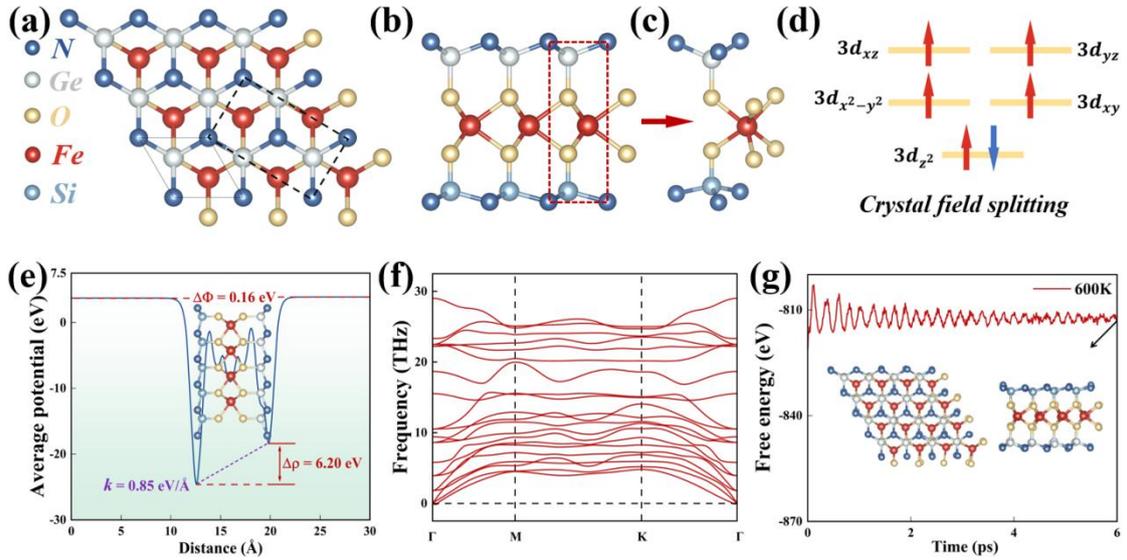

**Fig. 1** (a) Top and (b) side views of FeO$_2$SiGeN$_2$ monolayer. The primitive cell is marked by black solid line, and the orthorhombic supercell is marked by black dashed line for calculating piezoelectric coefficients. (c) The atomic structure of primitive cell which is marked by the red dashed line in (b). (d) Splitting of the Fe-3$d$ orbitals under the trigonal prismatic crystal field. (e) The planar average electrostatic potential energy of FeO$_2$SiGeN$_2$ monolayer and the slope of the purple dashed line corresponds to the intrinsic electric field. (f) The phonon dispersion spectrum of FeO$_2$SiGeN$_2$ monolayer. (g) The fluctuation of total energy and snapshots of geometric structures for FeO$_2$SiGeN$_2$ monolayer at 600K.

The designed Janus FeO$_2$SiGeN$_2$ monolayer (Figs. 1(a) and (b)) exhibits a hexagonal lattice with a $C_{3v}$ point group and same geometrical structures as the synthesized MoSi$_2$N$_4$ monolayer. As shown in Fig. 1(c), the Fe atom is sandwiched by O atoms to form a trigonal prism morphology similar to 2$H$-MoS$_2$, where Fe-3$d$ orbitals are split into $e_1$ ($d_{xy}$, $d_{x^2-y^2}$), $e_2$ ($d_{xz}$, $d_{yz}$) and $a$ ($d_{z^2}$), in which $a$ has the lowest energy and the electronic occuptation causes the magnetic moment of 4 $\mu_B$ in FeO$_2$SiGeN$_2$ monolayer (seen in Fig. 1(d)). The Janus structure composed of seven atomic layers in the N-Si-O-Fe-O-Ge-N stacked sequence indicates the breaking of inversion symmetry, which is reflected in the average electrostatic potential of FeO$_2$SiGeN$_2$ monolayer as shown in Fig. 1(e), where the values 6.20 and 0.16 eV of the drop of averaged potential $\Delta\rho$ and the electrostatic potential gradient $\Delta\Phi$, respectively, reflect the large out-of-plane dipole moment caused by the difference in electronegativity between Ge and Si atoms. The inherent electric field of 0.85 $eV/Å$ resulting from unequal charge transfer and out-of-plane dipole moment indicates the existence of the electric polarization and piezoelectric characteristics.

In order to determine the magnetic ground states of FeO$_2$SiGeN$_2$ monolayer in the range of 0-3 eV for $U_{\text{eff}}$, the supercell of 4×4×1 was used to study the energy difference of three types of antiferromagnetic (AFM) configurations, i.e. AFM1, AFM2 and AFM3 configurations (as shown in Supplementary Fig. S1), and the results were shown in Fig. S2. Although the energy difference between different configurations gradually decreases with the increase of $U_{\text{eff}}$, the FM state with the lowest energy is the most stable magnetic state. The lattice constant of FeO$_2$SiGeN$_2$ monolayer is 2.99 Å and the angle of Fe-O$_1$-Fe bond (89.47°) and Fe-O$_2$-Fe bond (89.13°) are close to 90°, implying that the indirect Fe-O-Fe FM super-exchange interaction leads to FM coupling between Fe atoms.[38-40] The energy difference between FM and different AFM configurations comes from the consequence of the competition between the indirect FM super-exchange and the direct AFM exchange. The indirect FM super-exchange interaction strength and direct AFM exchange interaction strength are proportional to $-\frac{t_{pd}^4 J_H^p}{(\Delta_{pd}+U_d)^4}$ and $\frac{t_{dd}^2}{U_d}$, respectively, where $t_{pd}$, $J_H^p$, $\Delta_{pd}$, $U_d$ and $t_{dd}^2$ representing the hybridization strength between Fe-3$d$ and O-2$p$ orbitals, the Hund's coupling strength of O-$p$ orbitals, the energy interval between Fe-3$d$ and O-2$p$ orbitals, the spin-splitting energy of Fe-3$d$ orbitals and the strength of direct hybridization between the Fe-3$d$ orbitals of the nearest-neighboring Fe atoms, respectively. The presence of $\Delta_{pd}$ and fourth power for $U_d$ makes a significant difference in the effect of $U_{\text{eff}}$ on the exchange interaction strength, i.e. the increased spin splitting reduces indirect FM super-exchange more rapidly than the direct AFM exchange.

To demonstrate stability of Janus FeO$_2$SiGeN$_2$ monolayer, the phonon spectra under -6, 0 and 6% biaxial strains are shown in Figs. 1(f) and S3, where the absence of negative frequency phonons in the whole Brillouin Zone (BZ) indicates the dynamic stability of the structure. The thermal stability of structure was investigated by performing AIMD simulations at 300 and 600K, and the small energy fluctuations and preservation of structural integrity proved that the excellent thermal stability of Janus FeO$_2$SiGeN$_2$ monolayer (in Figs. 1(g) and S4). Finally, we calculate the elastic constants $C_{11/22}$ (351.68 N/m) and $C_{12}$ (102.38 N/m), respectively, which meet the Born–Huang criteria (i.e. $C_{11/22} > C_{12} > 0$ and $C_{11}C_{22}-C_{12}^2 > 0$), proving the mechanical stability of the structure.

### 3.2. The origin of tunable MAE

MAE describing the magnetic anisotropy plays a crucial role in resisting the external noises induced by thermal and magnetic perturbations, and it is defined as $MAE = E_{[001]} - E_{[100]}$ by considering SOC in the Hamiltonian, where $E_{[001]}$ and $E_{[100]}$ represent the outside plane and inplane energy. The evolution of MAE with biaxial strains under different $U_{\text{eff}}$ is shown in Fig. S5. It can be found that with the application of strain, except $U_{\text{eff}} = 3.0$ eV, the system transits between perpendicular magnetic anisotropy (PMA) character and in-plane magnetic anisotropy (IMA), which represents the properties of XY ferromagnets, similar to the VSe$_2$ and LaBrI monolayer. In addition, the MAE as a function of angle $\theta$ can be express as MAE $(\theta) = K_1 cos^2\theta + K_2 cos^2\theta$ or MAE $(\theta) = K_3 sin^2\theta + K_4 sin^2\theta$, where $K_1(K_2)$ and $K_3(K_4)$ are the anisotropy constants and the angle $\theta$ takes the values of 0-180º. As shown in Figs. 2 (a), (b) and S6, the dependence of MAE on the polar angle ($\theta$) and the whole space for FeO$_2$SiGeN$_2$ monolayer at $U_{\text{eff}} = 0.0$ (2.0) eV exhibit easy magnetization z-axis ($xy$-plane) and MAE of -72 (514) $\mu eV$. Based on the significant regulatory effect of $U_{\text{eff}}$ and biaxial strain on the easy magnetization axis (plane), we take the evolution of MAE with $U_{\text{eff}}$ (in Fig. 2(c)) as an example to discuss the origin of MAE and its related regulatory mechanism.

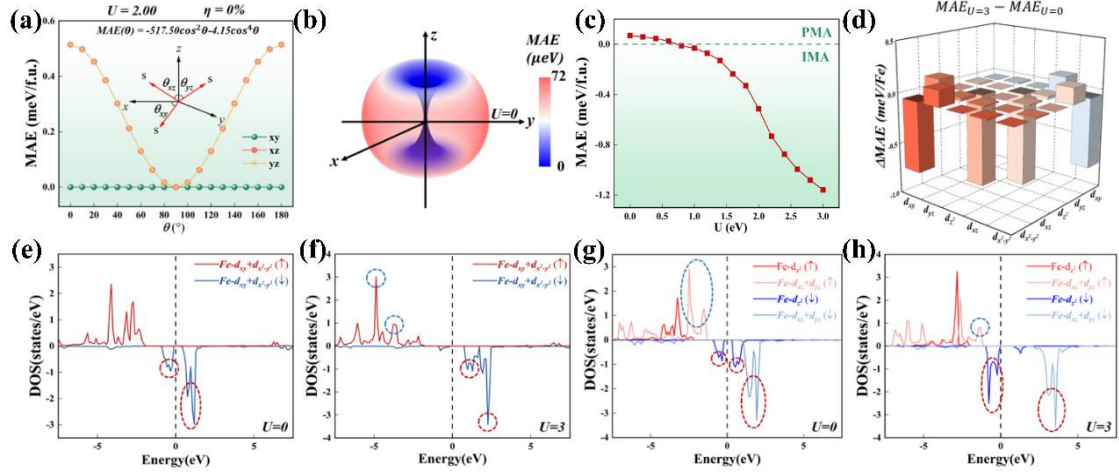

**Fig. 2** (a) Dependence of MAE on the polar angle ($\theta$) for the Janus FeO$_2$SiGeN$_2$ monolayer with the direction of magnetization lying on $xy$, $xz$ and $yz$ planes at $U_{\text{eff}} = 2$ eV. (b) Dependence of MAE for FeO$_2$SiGeN$_2$ monolayer on the whole space at $U_{\text{eff}} = 0$ eV. (c) The evolution of MAE in a unit cell with $U_{\text{eff}}$. (d) The difference in the orbital-resolved MAE of each Fe atom. The spin-polarized DOSs of the Fe-$d_{xy}+d_{x^2+y^2}$ orbitals at (e) $U_{\text{eff}} = 0$ eV and (f) $U_{\text{eff}} = 3$ eV. The spin-polarized DOSs of the Fe-$d_{z^2}$ and Fe-$d_{xz}+d_{yz}$ orbitals at (g) $U_{\text{eff}} = 0$ eV and (h) $U_{\text{eff}} = 3$ eV.

The MAE of the Janus FeO$_2$SiGeN$_2$ monolayer is contributed by Fe, O, Si, Ge and N atoms under the SOC effect. The element-resolved MAE as a function of $U_{\text{eff}}$ was calculated, and the $\Delta_{\text{MAE}}$ is defined as $\Delta_{\text{MAE}} = \text{MAE}_{U_{\text{eff}}=3/2/1\,eV} - \text{MAE}_{U_{\text{eff}}=0\,eV}$. It can be found that Fe atoms dominate the $\Delta_{\text{MAE}}$ of the system (seen in Figs. 2(d) and S7), and interestingly, the $\Delta_{\text{MAE}}$ from Fe and non-magnetic elements exhibit different behaviors: with the increase of $U_{\text{eff}}$, the contribution of non-magnetic atoms to $\Delta_{\text{MAE}}$ shows the upward trend; however, the contribution of Fe atoms to $\Delta_{\text{MAE}}$ shows the rapid decline trend.

According to the second-order perturbation, projected density of states (DOS) and coupling relationship, the orbital contribution of Fe-$3d$ orbitals with $U_{\text{eff}}$ is qualitatively discussed (in Fig. S8). Considering the occupied and unoccupied states near the Fermi level, MAE can be expressed as:[41,42]

$$MAE = \sum_{\sigma,\sigma'} (2\delta_{\sigma,\sigma'}-1)\xi^2 \times \sum_{o^\sigma,u^{\sigma'}} \frac{|\langle o^\sigma|\hat{L}_z|u^{\sigma'}\rangle|^2 - |\langle o^\sigma|\hat{L}_x|u^{\sigma'}\rangle|^2}{E_u^{\sigma'} - E_o^\sigma}, \quad (1)$$

where $\xi$, $\hat{L}_z$ and $\hat{L}_x$ are the strength of SOC and angular momentum operators, respectively. $|o^\sigma\rangle$ and $|u^{\sigma'}\rangle$ is the occupied and unoccupied state with spin $\sigma$ and $\sigma'$, whose energy is $E_o^\sigma$ and $E_u^{\sigma'}$. Combined with the dominance of Fe atoms on MAE, here we only consider the coupling of Fe-3d orbitals ($d_{xy}, d_{x^2-y^2}, d_{yz}$ and $d_{z^2}$) with the orbital angular momentum operator ($\hat{L}_z$ and $\hat{L}_x$). We first investigate the coupling relationship between Fe-($d_{xy}^\downarrow$-$d_{x^2-y^2}^\downarrow$) ($e_{1,u}^\downarrow$) and Fe-($d_{xy}^\downarrow$-$d_{x^2-y^2}^\downarrow$) ($e_{1,o}^\downarrow$) states in the case of $U_{\text{eff}} = 0$ eV (see Fig. 2(e)), which can be expressed as:

$$MAE_1 = \xi^2 \sum_{o,u} \frac{|\langle e_{1,u}^\downarrow|\hat{L}_z|e_{1,o}^\downarrow\rangle|^2 - |\langle e_{1,u}^\downarrow|\hat{L}_x|e_{1,o}^\downarrow\rangle|^2}{E_{e_{1,u}}^\downarrow - E_{e_{1,o}}^\downarrow}. \quad (2)$$

In the case of $U_{\text{eff}} = 3$ eV (see Fig. 2(f)), the disappearance of the occupied state $e_{1,o}^\downarrow$ near the Fermi level implies the emergence of the new coupling relationship between the $e_{1,o}^\uparrow$ state and $e_{1,u}^\downarrow$ state, which can be expressed as:

$$MAE_2 = -\xi^2 \sum_{o,u} \frac{|\langle e_{1,u}^\downarrow|\hat{L}_z|e_{1,o}^\uparrow\rangle|^2 - |\langle e_{1,u}^\downarrow|\hat{L}_x|e_{1,o}^\uparrow\rangle|^2}{E_{e_{1,u}}^\downarrow - E_{e_{1,o}}^\uparrow}. \quad (3)$$

Due to the coupling matrix elements between different d orbitals with the orbital angular momentum operator $\hat{L}_z$ and $\hat{L}_x$, the matrix elements $\langle e_{1,u}|\hat{L}_z|e_{1,o}\rangle$ and $\langle e_{1,u}|\hat{L}_x|e_{1,o}\rangle$ are 2i and 0, respectively. Apparently, the $MAE_1$ and $MAE_2$ can be simplified as: $MAE_1 = \xi^2 \sum_{o,u} \frac{4}{E_{e_{1,u}}^\downarrow - E_{e_{1,o}}^\downarrow}$ and $MAE_2 = -\xi^2 \sum_{o,u} \frac{4}{E_{e_{1,u}}^\downarrow - E_{e_{1,o}}^\uparrow}$, and the values of $MAE_1$ and $MAE_2$ are positive and negative respectively. The SOC interactions of two states with parallel and antiparallel spins dominate the MAE-Fe, respectively, which is reason for the negative transitions $d_{xy}$-$d_{x^2-y^2}$ in Fig. 2(d). In Fig. 2(g), the SOC interactions near the Fermi level are mainly contributed by $e_2$ ($d_{xz}, d_{yz}$) and $a$ ($d_{z^2}$), i.e., the coupling between Fe-$d_{z^2}^\downarrow$ ($a_u^\downarrow$) and Fe-$d_{xz}^\uparrow$-$d_{yz}^\uparrow$ ($e_{2,o}^\uparrow$), the coupling between Fe-$d_{z^2}^\downarrow$ ($a_o^\downarrow$) and Fe-$d_{xz}^\downarrow$-$d_{yz}^\downarrow$ ($e_{2,u}^\downarrow$) and the coupling between Fe-$d_{xz,u}^\downarrow$ and Fe-$d_{yz,o}^\uparrow$. Considering the matrix elements: $\langle a_u^\downarrow|\hat{L}_x|e_{2,o}^\uparrow\rangle = \sqrt{3}i$ and $\langle d_{xz,u}^\downarrow|\hat{L}_x|d_{yz,o}^\uparrow\rangle = i$, the $MAE_3$, $MAE_4$ and $MAE_5$ can be simplified as $MAE_3 = \xi^2 \sum_{o,u} \frac{3}{E_{a_u}^\downarrow - E_{e_{2,o}}^\uparrow}$, $MAE_4 = -\xi^2 \sum_{o,u} \frac{3}{E_{a_o}^\downarrow - E_{e_{2,u}}^\downarrow}$ and $MAE_5 = -\xi^2 \sum_{o,u} \frac{1}{E_{d_{xz,u}}^\downarrow - E_{d_{yz,o}}^\uparrow}$, respectively. The disappearance of Fe-$d_{z^2}^\downarrow$ ($a_u^\downarrow$) state at $U_{\text{eff}}$ = 3 eV (as shown in Fig. 2(h)) indicates the disappearance of the positive contribution $MAE_3$, i.e., the negative contribution of the transitions $e_2$ ($d_{xz}$, $d_{yz}$) and $a$ ($d_{z^2}$) to MAE increases with the $U_{\text{eff}}$. In conclusion, with increasing $U_{\text{eff}}$, the competition between states with parallel and antiparallel spins dominates $\Delta_{\text{MAE}}$, and the physical mechanism of MAE regulation by $U_{\text{eff}}$ or biaxial strain should be ascribed to the change of coupling between the states caused by the change of the relative position of the states.

### 3.3. Electronic properties and topological phase transitions

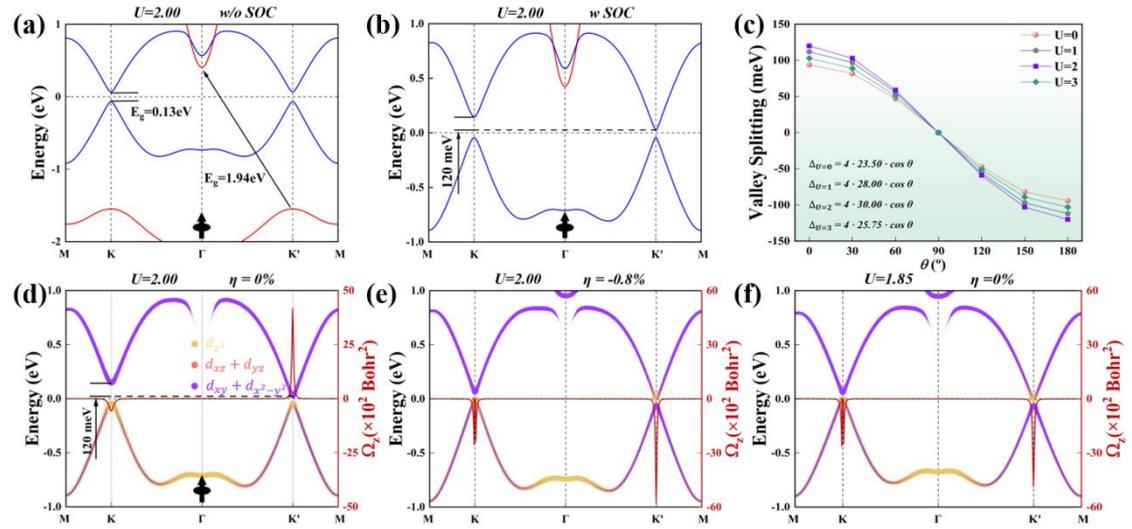

**Fig. 3** (a) The electronic band structures of Janus FeO$_2$SiGeN$_2$ monolayer (a) without (w/o) and (b) with (w) SOC at $U_{\text{eff}}$ = 2 eV. (c) Valley polarization as the function of magnetization direction under different $U_{\text{eff}}$, and the function is the corresponding cosine function. (d)-(f) The projected electronic band structure and Berry Curvature for FeO$_2$SiGeN$_2$ monolayer at different strain and $U_{\text{eff}}$.

In this section, we focus on the electronic band structure, which is crucial for spintronic (valleytronic) materials to manipulate spin (valley) degrees of freedom. When without the SOC effect, the valleys at valence band maximum (VBM) and conduction band minimum (CBM) exhibit the energy degeneracy and spin polarization (in Fig. 3(a)). Under the influence of $U_{eff}$ and biaxial strain, the transition between the bipolar ferromagnetic semiconductor and half-semiconductor can be found in FeO$_2$SiGeN$_2$ monolayer, which is crucial for generating 100% spin-polarized carriers in spintronics. In Fig. 3(b), when the SOC effect is considered, the energy degeneracy of valley at CBM is broken and a large valley polarization of 120 meV is generated, which is larger than most of ferrovalley materials such as LaBr$_2$ (33 meV) and VSe$_2$ (78.2 meV).[43,44] In Fig. S9, the valley polarization and spin-channels flip with the reversal of the magnetization direction of Fe atoms.

Next, we further analyze the intrinsic mechanism of the valley polarization generation and its evolutionary behavior. Based on the breaking of time-reversal symmetry, the interaction between the different spin-channels is ignored, and the SOC Hamiltonian can be expressed as:[45]

$$\widehat{H}_{SOC} \approx \widehat{H}_{SOC}^0 = \lambda \widehat{S}_{z'} \left( \widehat{L}_z \cos\theta + \frac{1}{2} \widehat{L}_+ + e^{-i\phi} \sin\theta + \frac{1}{2} \widehat{L}_- e^{+i\phi} \sin\theta \right), \quad (4)$$

where $\widehat{L}_{x,y,z}$ and $\widehat{S}$ represent the operator of orbital angular momentum and spin angular momentum, respectively. Both $\theta$ and $\phi$ are the magnetocrystalline angle, and $\widehat{L}_\pm = \widehat{L}_x \pm i\widehat{L}_y$. In Fig. 3(d), the basic functions can be chosen as: $|\psi_v\rangle = |d_{z^2}\rangle$, $|\psi_c^\tau\rangle = \frac{1}{\sqrt{2}}(|d_{x^2-y^2}\rangle + i\tau|d_{xy}\rangle)$, where $\tau = \pm 1$ indicates the valley index corresponding to the K and K' valleys, respectively. According to the representation of spherical harmonics $|L, L_z\rangle$ for atomic orbitals, $|3d_{x^2-y^2}\rangle = \frac{1}{\sqrt{2}}(|2, -2\rangle + |2, +2\rangle)$ and $|3d_{xy}\rangle = \frac{i}{\sqrt{2}}(|2, -2\rangle - |2, +2\rangle)$. Finally, we can obtain $E_c^K - E_c^{K'} = 4\alpha_{SOC} \cos\theta$, where $\alpha_{SOC}$ is the SOC-related constant. The variation trend of valley polarization with magnetocrystalline angle as cosine function is the same as the result of first principles, as shown in Fig. 3(c). With the application of $U_{eff}$ and biaxial strain, the values of valley polarization are exchanged between VBM and CBM, which can be interpreted as the exchange of band components between valleys (in Supplementary Figs. S10 and S11).

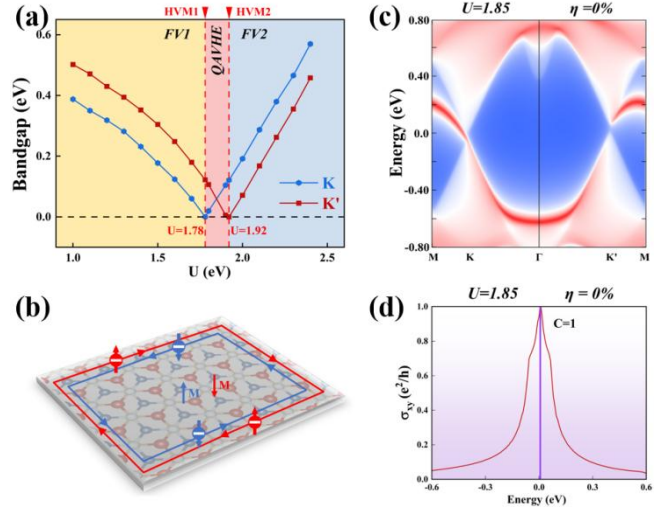

**Fig. 4** (a) The phase diagram and the bandgap of K and K' valley as a function of $U_{\text{eff}}$. (b) Schematic diagram of QAVHE in FeO2SiGeN2 monolayer. (c) The edge state and (d) anomalous Hall conductivity for FeO2SiGeN2 monolayer at $U_{\text{eff}} = 1.85$ eV.

In order to investigate the effect of the exchange of band components, Berry curvatures are calculated by Kubo formula:[46]

$$\Omega_{n,z}(k) = -\sum_{n\neq n'} \frac{2Im\langle\psi_{nk}|\hat{v}_x|\psi_{n'k}\rangle\langle\psi_{n'k}|\hat{v}_y|\psi_{nk}\rangle}{\left(E_n - E_{n'}\right)^2}, \qquad (5)$$

where $f_n$ represents the Fermi-Dirac distribution. $|\psi_{nk}\rangle$ and $\hat{v}_x$ ($\hat{v}_y$) are the Bloch wave function with eigenvalue $E_n$ and the velocity operator along x (y) direction, respectively. In Fig. 3(d), the Berry curvature of polarized-valleys (K and K' valleys) shows opposite signs and different absolute values, and under the action of an in-plane electric field, the carriers can obtain the anomalous transverse velocity ($v_\perp$): $v_\perp = \frac{1}{\hbar}\frac{\partial E_n(k)}{\partial k} + \dot{k} \times \Omega_z(k)$. As shown in Supplementary Figs. S11 and S12 with the application of $U_{\text{eff}}$ and biaxial strain, the polarized-valley have same valley features of initial band closure and then opening, accompanied by the exchange of the band components (i.e., the exchange of Fe-$a$ and Fe-$e_1$). Taking $U_{\text{eff}}$-induced topological phase transitions as an example, the bandgap of the K valley disappears at $U_{\text{eff}}$=1.78 eV, indicating that FeO2SiGeN2 monolayer enters the HVM state with 100% valley- and spin-polarization (in Fig. S11(b)).

When $U_{eff}$>1.78 eV, the reappearance of the K bandgap indicates that the HVM state disappears and the Berry curvature of K valley reverses with the exchange of the band components, which is the reason for the exchange of valley polarization values between VBM and CBM. Meanwhile, the Berry curvature of the same sign at polarized-valleys indicates that the system has entered the QAVH phase from FV1, which can be proved by single chiral edge state (CES), quantized AHC ($\sigma_{xy} = Ce^2/h$) and Chern number of $C = 1$ (in Fig. 4). As the $U_{eff}$ increases to 1.92 eV, the system again enters the HVM state, which is opposite to the HVM at $U_{eff}$=1.78 eV. The system enters the FV state again when $U_{eff}$>1.92 eV, however, the Berry curvature of FV2 is reversed in relation to FV1 by two exchanges of band components. The effect of biaxial strain on topological phase transitions is similar to that of $U_{eff}$, and the related topological phase transition diagram are displayed in Fig. S12.

### 3.4. The two-band strained k·p model

To understand the topological phase transition mechanism in FeO2SiGeN2 monolayer, a two-band strained ***k·p*** model with SOC, magnetic interactions and strain is constructed. According to the orbital projection analysis in Fig. 3(d), the basic functions can be chosen as: $|\psi_v\rangle = |d_{z^2}\rangle$, $|\psi_c^\tau\rangle = \frac{1}{\sqrt{2}}(|d_{x^2-y^2}\rangle + i\tau|d_{xy}\rangle)$. The total Hamiltonian with SOC includes three parts, i.e. $H_k = H_0 + H_{SOC} + H_{ex}$, which correspond to the orbital interactions of the nearest-neighbor hopping $H_0$, the influence originating from the SOC effect $H_{SOC}$ and inherent exchange interaction of Fe-3*d* electrons $H_{ex}$. The three terms are given by:[47-49]

$$H_0 = \begin{bmatrix} (f_0 - \frac{f_1}{2}) - (f_2 - f_3)(\varepsilon_{xx} + \varepsilon_{yy}) & ta(\tau k_x + ik_y) - f_4(\varepsilon_{xx} - \varepsilon_{yy} - 2i\varepsilon_{xy}) \\ ta(\tau k_x + ik_y) - f_4(\varepsilon_{xx} - \varepsilon_{yy} - 2i\varepsilon_{xy}) & (f_0 - \frac{f_1}{2}) - (f_2 - f_3)(\varepsilon_{xx} + \varepsilon_{yy}) \end{bmatrix}, \quad (6)$$

$$H_{SOC} = \begin{bmatrix} \tau s\lambda_c & 0 \\ 0 & \tau s\lambda_v \end{bmatrix}, \quad (7)$$

$$H_{ex} = \begin{bmatrix} -sM_c & 0 \\ 0 & -sM_v \end{bmatrix}, \tag{8}$$

where, $f_0$, $f_1$, $f_2$ and $f_3$ are band-gap adjustment parameters. $f_4$ is the pseudo gauge-field term, $\varepsilon_{ij}$ is the element of the strain tensor and $t$ is the effective nearest-neighbor hopping integral. Spin is indexed by $s$ with +1 (-1) for spin-up (spin-down) channel. When biaxial strain is applied, $\varepsilon_{xx} = \varepsilon_{yy}$ and $\varepsilon_{xy} = 0$, and consider the spin index ($s = \pm 1$) near the Fermi level, the direct bandgaps of K and K' valleys can be obtained by diagonalizing $H_k$:

$$E_g^{+1} = |f_1 + \tau(\lambda_v - \lambda_c) + (M_c - M_v) - 4f_3\varepsilon_{xx}|, \tag{9}$$

$$E_g^{-1} = |f_1 + \tau(\lambda_c - \lambda_v) + (M_v - M_c) - 4f_3\varepsilon_{xx}|. \tag{10}$$

Obviously, the SOC effect and inherent exchange interaction have opposite effects on the direct bandgaps of K and K' valleys under different spins, and the presence of valley index $\tau$ in $\tau(\lambda_v - \lambda_c)$ term indicates that SOC effect has opposite effect in polarized-valleys. In the absence of SOC effect, the bandgap of the K and K' valleys both are $|f_1 + (M_c - M_v) - 4f_3\varepsilon_{xx}|$, which implies the energy degeneration between the polarized-valleys. The $4f_3\varepsilon_{xx}$ term associated with biaxial strain is the basis for the evolution of the bandgaps, at the same time, the variation trend of band gap at K and K' valleys verifies the correctness of this parameter. The induction mechanism of topological phase transitions with $U_{eff}$ can be understood by analogy with biaxial strain.

### 3.5 The Piezoelectric AVHE mechanism

Based on the unique Janus structure of FeO$_2$SiGeN$_2$ monolayer, the uniaxial strain can induce both in-plane and out-of-plane piezoelectric response ($e_{11}/d_{11} \neq 0$ and $e_{31}/d_{31} \neq 0$), while biaxial strain can induce only out-of-plane piezoelectric response. Using the Voigt symbol, the piezoelectric tensor $e_{ik}$ and piezoelectric coefficient $d_{ij}$ for $P3m1$ symmetry can be expressed as:[50]

$$e_{ik} = \begin{pmatrix} e_{11} & -e_{11} & 0 \\ 0 & 0 & -e_{11} \\ e_{31} & e_{31} & 0 \end{pmatrix} \text{ and } d_{ij} = \begin{pmatrix} d_{11} & -d_{11} & 0 \\ 0 & 0 & -2d_{11} \\ d_{31} & d_{31} & 0 \end{pmatrix}. \tag{11}$$

Meanwhile, $d_{11}$ and $d_{31}$ can be express by $e_{ik} = d_{ij}C_{jk}$: $d_{11} = \frac{e_{11}}{C_{11}-C_{12}}$ and $d_{31} = \frac{e_{31}}{C_{11}+C_{12}}$.

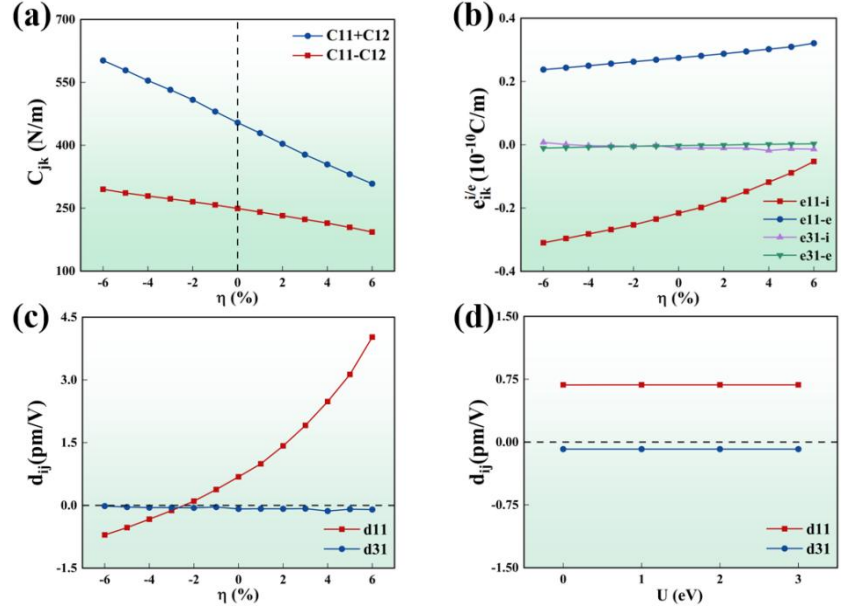

**Fig. 5** (a) The elastic constants $C_{jk}$ as a function of biaxial strain. (b) The contributions of electrons and ions for piezoelectric tensor $e_{ik}$ as a function of biaxial strain. The piezoelectric coefficient $d_{ij}$ as a function of (c) biaxial strain and (d) $U_{\text{eff}}$.

As shown in Figs. 5(a)-(c), $C_{11} \pm C_{12}$, $e_{ik}$ of electronic and ionic contribution and $d_{ij}$ with different biaxial strain are shown by using the orthorhombic supercell (see Fig. 1(a)). With the application of biaxial strain (-6 to 6%), the ionic contribution to $e_{11}$ decreases while the electronic contribution for $e_{11}$ increases, resulting in the in-plane piezoelectric response changing from -0.71 to 4.03 pm/V. Due to the small contributions of electrons and ions to $e_{31}$, the system has a negligible out-of-plane piezoelectric response (-0.04 — -0.13 pm/V), which should be attributed to the tiny electronegativity difference between Si and Ge atoms. In Fig. 5(d), unlike the regulation of $d_{11}$ by biaxial strain, the piezoelectric coefficients $d_{ij}$ is robust against $U_{\text{eff}}$ for $FeO_2SiGeN_2$ monolayer. As shown in Supplementary Fig. S13, the polar coordinates can be used to describe the Young's modulus $Y$ and Poisson's ratio $v$ of 2D materials:

$$Y(\theta) = \frac{C_{11}C_{22}-C_{12}^2}{C_{11}\sin^4\theta+(\frac{C_{11}C_{22}-C_{12}^2}{C_{66}}-2C_{12})\sin^2\theta\cos^2\theta+C_{22}\cos^4\theta)}, \quad (12)$$

$$\nu(\theta) = \frac{C_{12}\sin^4\theta-\left(C_{11}+C_{22}-\frac{C_{11}C_{22}-C_{12}^2}{C_{66}}\right)\sin^2\theta\cos^2\theta+C_{12}\cos^4\theta}{C_{11}\sin^4\theta+\left(\frac{C_{11}C_{22}-C_{12}^2}{C_{66}}-2C_{12}\right)\sin^2\theta\cos^2\theta+C_{22}\cos^4\theta}. \quad (13)$$

Here, $\theta$ represents the angle relative to the x-axis. The $Y$ of $FeO_2SiGeN_2$ exhibits the isotropic characteristics in $xy$ plane, and shows a change from 213 N/m to 355 N/m, which reflects the lower stiffness and greater flexibility of system compared to graphene (350 N/m). Meanwhile, the Poisson's ratio of $FeO_2SiGeN_2$ is higher than that of graphene (0.175) and H-BN (0.211).[51]

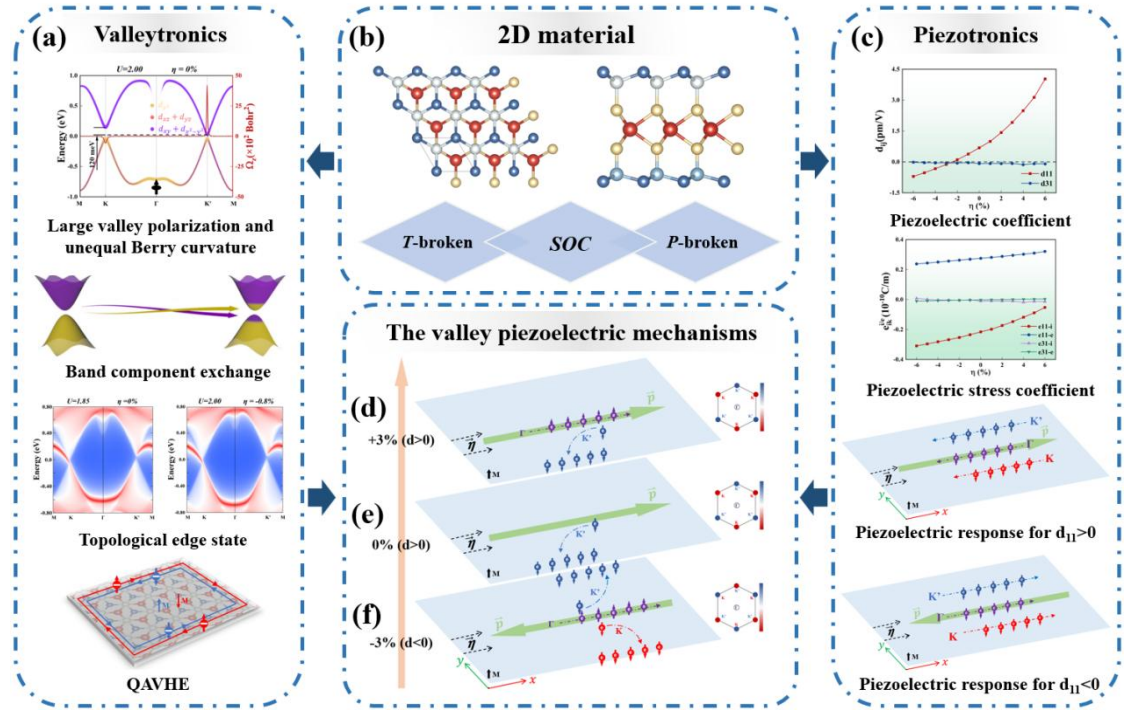

Fig. 6 The valley piezoelectric mechanisms. (a) The large valley polarization, band component exchange, topological edge state and QAVHE. (b) 2D materials with *T*-broken and *P*-broken. (c) The piezoelectric coefficient, piezoelectric stress coefficient and related piezoelectric response. (d)-(e) The piezoelectric-AVHE with piezoelectric polarization driving carrier.

The introduction of Berry curvature, which can be regarded as the pseudomagnetic field, allows the emergence of AVHE without dependence on the external magnetic field (in Fig. 6(a)). Considering the coexistence of valley and piezoelectric polarization under *P*-broken, we propose a piezoelectric-AVHE which is independent of the external field in ferrovalley system, where the external in-plane electric field is replaced by the intrinsic polarized electric field generated by piezoelectric polarization (in Fig. 6(b)). The intrinsic polarized electric field induced by +0.1% uniaxial strain is discussed by taking $FeO_2SiGeN_2$ monolayer in the absence of biaxial strain as an example. Based on the same permittivity in the x and y directions, the in-plane permittivity of $FeO_2SiGeN_2$ is $\varepsilon_{xx} = \varepsilon_{yy} = 8.06$. The in-plane piezoelectric response $d_{11}$ and Young's modulus $Y$ under 0% biaxial strain are 0.68 pm/V and 289 N/m, respectively, and the corresponding stress $\sigma$ of uniaxial x-direction strain can be expressed as: $\sigma = Y \cdot \varepsilon$. The direction of stress-induced piezoelectric polarization is closely related to $d_{11}$ ($P = d_{11} \cdot \sigma$), i.e. the direction of piezoelectric polarization is the same (opposite) with the direction of stress $\sigma$ when $d_{11} > 0$ ($d_{11} < 0$) (in Fig. 6(c)). The intrinsic polarized electric field $E$ can be expressed as: $E = \frac{P}{\varepsilon_{xx} \cdot \varepsilon_0}$, where $\varepsilon_0$ is vacuum permittivity. It can be found that +0.1% uniaxial strain can induce the intrinsic polarized electric field of $0.39 \times 10^7$ V/m in $FeO_2SiGeN_2$. Such a large in-plane polarized electric field induced by 0.1% uniaxial strain indicates the great potential of piezoelectric polarization as a means of driving carrier motion.

In Fig. 6(a), the large valley polarization and clean energy window makes it possible for piezoelectric response to excite carriers in polarized-valleys, but this discussion is beyond the scope of this work. By proper charge doping to assist the intrinsic polarization field, the carriers of the polarized-valleys obtain the anomalous transverse velocity $v_\perp$ under the unequal Berry curvature (in Figs. 6(d)-(f)). The presence of the polarized electric field caused by piezoelectric polarization and intrinsic ferromagnetism make the realization of piezoelectric-AVHE independent of external field, and the strain-induced symbolic inversion of Berry curvature and $d_{11}$ provides new regulatory ideas into valley Hall-related effect.

## 4. Conclusions

In summary, based on first-principles calculations, we have determined that Janus $FeO_2SiGeN_2$ monolayer is a ferrovalley material with great stability and large valley polarization up to 120 meV. Due to the significant $U_{eff}$ influence on Fe-3$d$ orbitals, based on the perturbation theory, we reveal that the tunable MAE of this system result from competition and coupling of orbital-resolved MAE contributed by Fe atoms. The topological phase transitions of FeO$_2$SiGeN$_2$ monolayer can be induced by biaxial strain and $U_{eff}$, and a two-band strained $\boldsymbol{k} \cdot \boldsymbol{p}$ model has been established to understand the key role of strain and SOC in phase transitions. Based on AVHE, we propose the concept of piezoelectric-AVHE, in which the carrier is driven by the intrinsic polarized electric field generated by the piezoelectric polarization rather than the external in-plane electric field. We integrate the transport properties involving charge, spin and valley degrees of freedom into a single system, i.e. piezoelectric transport, AVHE, piezoelectric-AVHE and QAVHE, which could provide a blueprint for exploring valley Hall-related transport.

## Conflicts of interest

All authors declare that we have no financial and personal relationships with other people or organizations that can inappropriately influence our work

## Acknowledgments

We would like to acknowledge the support from the Natural Science Foundation of Hebei Province (No. A2020202010).